# Magnetic flux trapping in porous superconductors

D. M. Gokhfeld

Kirensky Institute of Physics, Krasnoyarsk Scientific Center, Siberian Branch, Russian Academy of Sciences, Krasnoyarsk, 660036 Russia.

The magnetization of superconducting samples is influenced by their porosity. In addition to structural modifications and improved cooling, the presence of pores also plays a role in trapping magnetic flux. Pores have an impact on the irreversibility field, the full penetration field, and the remnant magnetization. Generally, as porosity increases, these parameters tend to decrease. However, in the case of mesoscopic samples or samples with low critical current densities, increased porosity can actually enhance the trapping of magnetic flux.

## 1. Introduction

An effective cooling is crucial for many applications of superconductors. In the case of overheating, stable functioning is hindered and the superconductor can be even damaged [1, 2]. Improvement of cooling can be achieved by perforation providing a coolant to infiltrate the sample and to withdraw the emitted heat. At the same time perforation can increase oxygenation and prevent a cracking during the sample synthesis. It is believed that the same positive effects for superconductors can be achieved due to porosity.

Porous superconductors are the special class of novel materials [3–5]. They have properties inherent both bulk 3-D and quasi 2-D systems [6, 7] because pores are defects having high specific surface. Pinning of vortices and the critical current density depend on the concentration, form, and size of defects. Earlier works concerning pinning were focused on defects of about the superconductor coherence length [8, 9]. As it was believed, the larger defects should depress the critical current due to distortions of current trajectories [10]. At the same time, the coolant percolation needs the infinite pore cluster and the pore diameter of about 0.1-1 mm [11]. Porous high-$T_c$ superconductors are realized in forms of foam, fiber, fabric, and polycrystal [5]. Superconducting foams investigated since 2002 [12] have high porosity (up 80 %) and 1 mm pores. For such samples, high porosity provides chemical uniformity and effective cryocooling.

Describing these systems is complicated [5, 13] because trajectories of currents and vortex configurations in porous superconductors are tangled [14, 15]. The simpler concerning model considered in [16–18] is a long perforated cylinder. This model accounts that holes in the sample trap an excess magnetic flux due to their surface barriers [19]. The corresponding entrance field for magnetic vortices is about the thermodynamical critical field of the superconductor [20]. It was found that the perforated sample with a uniform hole spreading and the sample with a single central hole have an analogous dependence of the trapped magnetic flux on an effective hole area [17, 18].

The presented work focuses only on the effect of the pores on the trapped magnetic flux. An optimal porosity is expected in an analogy to the optimal hole area in the perforated superconductors [17, 18]. The aim is estimating this optimal porosity. The effects of porosity on oxygenation, cracking, and cooling, which can be more important for superconducting performance, are not accounted here.

## 2. Model

Let us firstly consider a perforated sample, which is a long superconducting cylinder with diameter $D \gg \lambda$, where $\lambda$ is the London penetration length. The external magnetic field $H$ is applied parallel to the major axis of the cylinder, so the demagnetization is neglected. The maximum value $H_{max}$ of $H$ is higher than the full penetration field $H_p$. The sample contains identical holes drilled parallel to the major axis and uniformly distributed over the end face of the cylinder. Material-related parameters (the area of the sample end face, $S$; the flux trapped in the unperforated sample, $\Phi_{pin0}$; the corresponding remnant magnetic field, $B_{pin0} = \Phi_{pin0}/S$) and hole-related parameters (the coefficient depending on an arrangement of the holes, $A_h$; the number of the holes, $N_h$; the diameter of the hole, $D_h$; the average magnetic field in the holes, $B_h$) characterize the sample.

Perforations decrease the useful area and the number of inner pinning centers. The flux $\Phi_{pin}$ trapped by intrinsic pinning centers decreases as $\Phi_{pin} \approx \Phi_{pin0}(1 - n_h)$, here $n_h$ is a perforation coefficient. The perforation coefficient is expressed as $n_h = \pi A_h N_h D_h^2/(4S)$. It is suggested that $A_h = 1$ in the case of uniform arrangement. Concentrating the holes in a circle segment and curving the current trajectories due to holes [16] can increase $A_h$.

The trapped magnetic flux depends on $n_h$ as

$$\Phi \approx \Phi_{pin0}(1 - n_h)(1 + n_h k_h), \qquad (1)$$

where $k_h$ is the reduced excess field in the holes, $k_h = B_h/B_{pin0}$. This expression has been used to describe the trapped magnetic flux of perforated samples simulated by the Monte-Carlo method [18].

Now consider a superconductor having nearly spherical pores with different diameters. The cross-section of a porous cylinder by the plane perpendicular to its major axis looks like the same cross-section of the perforated sample. Unlike the perforated sample, voids corresponding to the cross-section of the pores have different sizes, and an arrangement of the voids depends on the position of the cross-section plane. As well, the magnetic field distribution depends on the coordinate z. However, the trapped flux is independent of z value. It should be also noted that the trajectories of currents flowing through the porous superconductors are not confined to planes in the sample, but can be three-dimensional [15]. Due to elasticity of superconducting vortex lines [9, 21], each line can warp and skewer many pores to minimize the vortex energy [22, 23]. We suggest to consider one cross-section characterizing by the area $S$, the density of voids $\varphi_v$, and the averaged diameter of the voids $D_v$. The value of $D_v$ is smaller than the average pore diameter $D_p$, the simple math gives $D_v = \pi D_p/4$.

The correspondence between perforated and porous samples allows us to express the trapped flux for porous media:

$$\Phi \approx \Phi_{pin0}(1-\varphi)(1+k_h\varphi), \qquad (2)$$

with $\varphi = \pi^3 N_p D_p^2/(64S)$. Here it was taken into account that $\varphi_v = \varphi$ in the considered case of porous media with random structure [24]. The three important parameters $\varphi_{lim}$, $\varphi_{max}$, and $\Phi_{max}$ are resulted from (2). The trapped flux is greater than $\Phi_{pin0}$ when $\varphi < \varphi_{lim}$, where

$$\varphi_{lim} = 1 - \frac{1}{k_h}. \qquad (3)$$

The maximum trapped flux is achieved for $\varphi = \varphi_{max}$, where

$$\varphi_{max} = \frac{1}{2}\left(1 - \frac{1}{k_h}\right). \qquad (4)$$

The corresponding maximum value of trapped flux is

$$\Phi_{max} = \frac{\Phi_{pin0}}{4}\left(2 + k_h + \frac{1}{k_h}\right). \qquad (5)$$

The value of $k_h$ for pores is assumed to be the same as for holes [19]: $k_h \approx [(B_s/B_{pin0})^2+1]^{1/2}$, where $B_s = \Phi_0/(4\pi\lambda\xi)$, $\xi$ is the coherence length of the superconductor, and $\Phi_0 = 2.07*10^{-15}$ Wb.

### 3. Discussion

The trapped magnetic flux $\Phi$ depends on the critical current density, the sample size, and the porosity. The dependence of $\Phi$ on $\varphi$ for a cylindrical YBCO sample with $D = 2$ cm is demonstrated in Fig. 1a. Using $B_{pin0} \approx \mu_0 j_c D/6$ for the considered cylinder geometry and given typical parameters of large-grain bulk REBCO samples ($j_c = 10^8$ A/m$^2$ and $B_s = 0.17$ T at 77 K), one obtains $k_h \approx 1.08$ for $D = 2$ cm. The trapped magnetic flux is increased by adding some porosity in the sample until $\varphi \approx 0.036$ (see inset in Fig. 1a). The increase of $\Phi$ is negligible in this case. As the porosity increases, the trapped flux decreases. The same decrease was observed in resent study [25]. The porosity of investigated YBCO foam samples is equal to 0.75 [15, 26] (arrow in Fig. 1a). It is much higher than the optimal value.

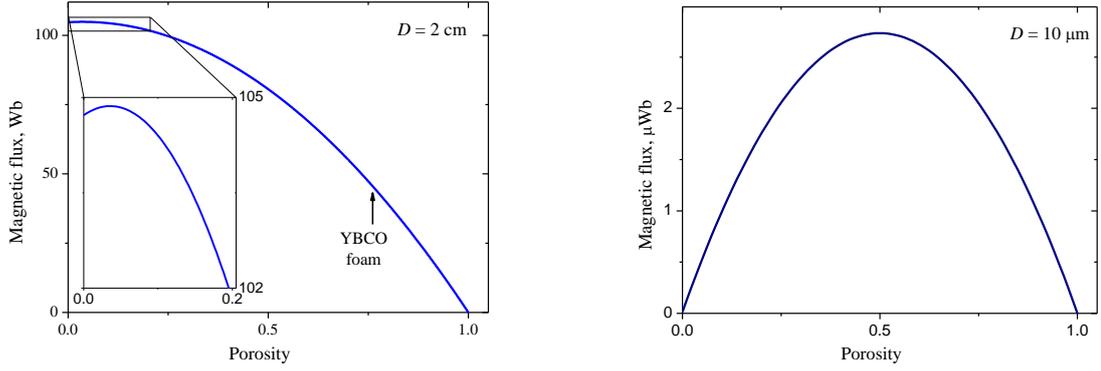

Fig. 1. Effect of porosity φ on the trapped magnetic flux Φ for a 2 cm sample (a) and a 10 μm sample (b). Arrow indicates porosity of YBCO foam. Insert (a) demonstrates the Φ(φ) dependence for small values of φ.

The low value of $k_h$ means that the flux trapping on the pores is negligible in the large singe grain superconductors. In addition, thermal fluctuations can further suppress the surface barrier in high-$T_c$ superconductors [27].

The higher values of $k_h$ can be realized in the porous superconductors only with the smaller sizes or the smaller values of critical current density. Fig. 1b demonstrates the Φ(φ) dependence for a cylindrical YBCO sample with $D$ = 10 μm. For this sample, $k_h \approx 833$ and the increase of Φ is significant. It is seen that the maximum position shifts to greater φ and the peak corresponds to φ ≈ 0.499.

Figure 2 shows the effect of the sample diameter on the maximal possible increase of trapped magnetic flux $\Phi_{max}/\Phi_{pin0}$. As it is seen in Fig. 2, mesoscopic samples can significantly increase Φ due to the porosity. Apparently, some unaccounted mechanisms may hinder this huge increase of the trapped magnetic flux.

Consider potential effects of porosity on the sample magnetization. Pores can facilitate a movement of magnetic flux into the sample inner and accordingly decrease the full penetration field and the magnetization width Δ$M$. The surface equilibrium magnetization depends on the specific surface. The specific surface increases with porosity. It increases the surface equilibrium magnetization and decreases the bulk non-equilibrium magnetization of the porous superconductor [28, 29]. Thus, increasing porosity makes the magnetization hysteresis loops more asymmetric along the $H$ axis and decreases the irreversibility field (Fig. 3). When (3) is satisfied, the remnant magnetization is increased due to the excess flux trapped in the pores. This effect is awaited only for mesoscopic samples. If (3) is not satisfied, the remnant magnetization will decrease.

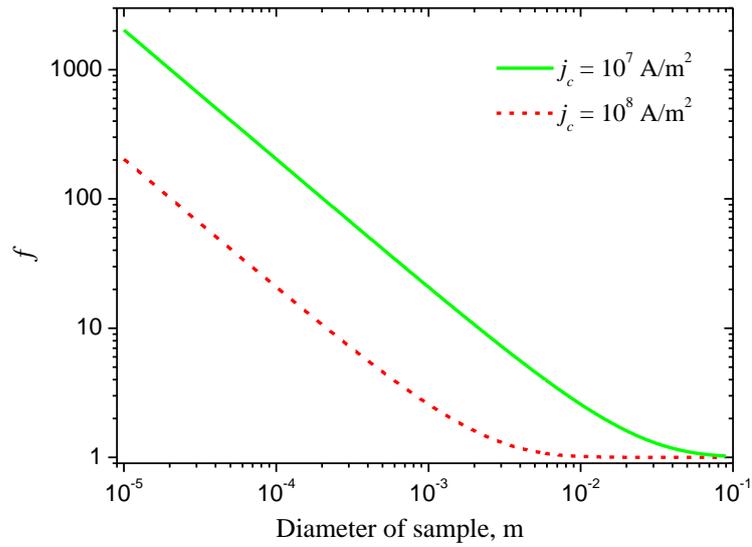

Fig. 2. Potential improvement of trapped magnetic flux.

It should be noted that in the cases of polycrystalline superconductors and fabrics, which are two-level superconductors [30], pores trap the excess magnetic flux only in small fields till 100 Oe. At the higher fields the most magnetic flux is trapped by the defects into granules and crystallites.

The significant effect of pore on the magnetization was found in opal like low-$T_c$ superconductors. These materials can be as I-type, so II-type superconductors [7]. Tuning the pore size can transform the type of superconductivity. The most magnetic flux in these materials is trapped by superconducting contours around pores [6, 31].

Possible room-temperature superconductors will have low values of $j_c$. We believe that be these materials can be significantly facilitated by porosity. Porosity can increase the remnant magnetization of these forthcoming materials and simultaneously lighten the sample weight.

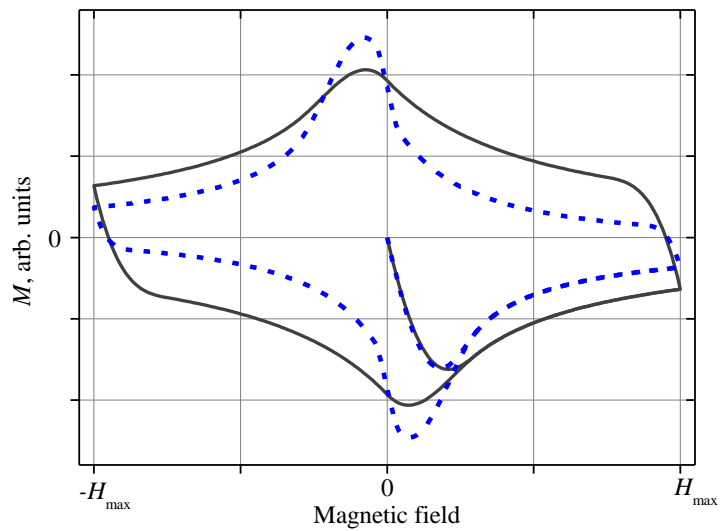

Fig. 3. The hysteresis magnetization loops of pore-free superconductor (solid line) and porous superconductor (dashed line).

**Conclusion**

Porosity affects the magnetic properties of superconducting samples. The loop asymmetry is increased and the irreversibility field is decreased by the increased porosity. It is found that porosity generally decreases the trapped magnetic flux in macroscopic samples. The advantageous effect of porosity on the trapped flux is awaited only for small-size or low-$j_c$ superconductors that are fulfilling the condition (3).